# Machine Learning Reveals the Seismic Signature of Eruptive Behavior at Piton de la Fournaise Volcano

C. X. Ren[1,2], A. Peltier [3,4], V. Ferrazzini [3,4], B. Rouet-Leduc[2], , P. A. Johnson[4], F. Brenguier[5]

[1]Space Data Science and Systems Group, Los Alamos National Laboratory, MS D440, Los Alamos, New Mexico 87545, USA

[2] Geophysics Group, Los Alamos National Laboratory, MS D446, Los Alamos, New Mexico 87545, USA

[3] Université de Paris, Institut de physique du globe de Paris, CNRS, F-75005, Paris, France

[4]Observatoire volcanologique du Piton de la Fournaise, Institut de physique du globe de Paris, F-97418 La Plaine des Cafres, France

[5] ISterre, Université Grenoble Alpes, Gières, France

Corresponding author: Christopher Ren (cren@lanl.gov)

**Key Points:**

- Statistical features derived from time-windowed, filter banked seismic data can be an effective way to characterize eruptive behavior of volcanoes.

- Supervised learning allows us to determine the eruptive state of the volcano given a single time window of raw seismic data from a single station.

- Spectral clustering can reveal different phases of eruptions and differences between various eruptions.

**Abstract**

Volcanic tremor is key to our understanding of active magmatic systems but, due to its complexity, there is still a debate concerning its origins and how it can be used to characterize eruptive dynamics. In this study we leverage machine learning (ML) techniques using 6 years of continuous seismic data from the Piton de la Fournaise volcano (La Réunion island) to describe specific patterns of seismic signals recorded during eruptions. These results unveil what we interpret as signals associated with various eruptive dynamics of the volcano, including the effusion of a large volume of lava during the August-October 2015 eruption, as well as the closing of the eruptive vent during the September-November 2018 eruption. The ML workflow we describe can easily be applied to other active volcanoes, potentially leading to an enhanced understanding of the temporal and spatial evolution of volcanic eruptions.

**Plain Language Summary**

A good understanding of volcanic activity is key to managing volcanic hazards resulting from eruptive activity. Volcanic tremor is a continuous seismic signal often seen during eruptions associated with the flow of magma through the volcano, and is thus an extremely useful tool in characterizing the progression and phases of eruptions. In this study we study this signal at the Piton de la Fournaise volcano, on La Réunion island. Using machine learning (ML) algorithms we investigate characteristics of this signal emitted by the volcano during eruptions to reveal the fundamental frequency at which it occurs, as well as changes in eruptive state which occur during some eruptions in our dataset. This workflow may be applied to other volcanos to further our understanding of eruptive dynamics.

**1 Introduction**

Long lasting seismic signals known as volcanic tremors are almost ubiquitously present in eruptive episodes at volcanoes **(Jellinek & Bercovici, 2011)**. These seismic signals are thought to be critical in characterizing the magma migration pathways in the internal plumbing system of volcanoes, as they are typically linked with magma propagation **(B. A. Chouet, 1996)** as well as degassing from eruptive vents **(Battaglia, Aki, & Staudacher, 2005)**, and are thus of interest in the context of both monitoring and forecasting volcanic activity . Attempts to forecast and monitor volcanic eruptions must take into account that the style and magnitudes of eruptions vary drastically, even throughout the duration of a single eruption **(Battaglia, Aki, & Staudacher, 2005; Chardot et al., 2015; Kurokawa et al., 2016)**. Despite the inherent complexity of volcanic systems, there have been many demonstrations of the utility of volcanic tremor in monitoring eruptions. **Battaglia & Aki (2003)** demonstrated the eruptive tremor sources at Piton de la Fournaise volcano were good indicators of the locations of eruptive fissures, in a subsequent study **Battaglia, Aki, & Ferrazzini, (2005)** demonstrated that the cumulative amplitude of tremor recorded throughout an eruption could be used to estimate the volume of lava erupted. Several studies have also shown that analyzing the spectral content and location of volcanic tremor can help track the evolution of  magma migration **(Di Lieto et al., 2007; Jellinek & Bercovici, 2011; Kurokawa et al., 2016)**

The application of machine learning (ML) techniques to the analysis of geophysical signals has become widespread in diverse settings such as the analysis of laboratory experiments **(Hulbert et al., 2018; Rouet-Leduc et al., 2017; Rouet-Leduc, Hulbert, Bolton, et al., 2018)**, tracking slow-slip in real Earth **(Rouet-Leduc, Hulbert, & Johnson, 2018)**, phase-association for the development of earthquake catalogs to name a few **(McBrearty et al., 2019; Ross et al., 2019)**.

With regards to the applications of ML to study and characterization of volcanoes, the primary applications thus far have been in the classification of volcano-seismic signals (**Hibert et al., 2017; Malfante et al., 2018; Titos et al., 2019**). In this work, we describe how statistical features derived from the continuous seismic signal recorded at Piton de la Fournaise volcano can be utilized to build machine learning models which reveal the characteristic eruptive tremor and eruptive dynamics of volcanic eruptions.

**1.1 Regional Setting**

Piton de la Fournaise is an active volcano situated on La Réunion, a hot-spot basaltic island in the western part of the Indian Ocean located approximately 800 km east of Madagascar. It is one of the most active volcanoes in the world, exhibiting 71 eruptions between 1985 and 2018 (**Duputel et al., 2019; Roult et al., 2012**).

Recent activity has been focused in the Enclos Fouqué caldera, formed about 5000-3000 years ago (**Ort et al., 2016**), inside which a terminal cone was formed as a consequence of frequent effusive eruptions characterized by lava fountains and lava flow emissions. Following 41 months of rest between end of 2010 and June 2014, eruptive activity at Piton de la Fournaise renewed on June 20, 2014. Between 2014 and March 2019: 14 eruptions occurred on the flank or at the base of the terminal cone. The locations of the fissures generated by these eruptions are shown in Figure 6. Eruptions are fed by a magma plumbing system consisting of a succession of reservoirs spreading from depth greater than to 30 km below the external western flank of the volcano

**(Michon et al., 2015)** to 2 km below the summit craters **(Di Muro et al., 2014; Peltier et al., 2009)**, where most eruptions initiate.

## 2.1 Seismic Data

For this study we leverage 6 years of continuous seismic data recorded across the Observatoire Volcanologique du Piton de la Fournaise (OVPF) network shown in Figure 1. The network consists of stations equipped with short-period seismometers and broadband seismometers, recording at a sampling rate of 100Hz. The data shown in this manuscript are primarily from the seismic station of the Cratere Bory (BOR) site although the analysis described in sections 2.2 and 2.3 has been performed for most of the seismic stations situated in the the Enclos Fouqué belonging to the OVPF network caldera we find that data recorded at the BOR site provides the best performance for our eruptive state classifier. We note that in the period under study at BOR seismic data for the April 4$^{th}$ eruption is missing, and thus this eruption does not appear in our dataset We also use data from the Enclos Sery Sud (CSS), Chateau Fort (FOR) and the Faujas (FJS) sites to characterize signals recorded throughout the eruptions, as these stations are approximately equidistant from the Dolomieu crater **(Plaen et al., 2016)** and have a good amount of overlap in terms of available data during the periods of interest.

## 2.2 Feature Building

In order to reduce the continuous seismic signal recorded at various stations at Piton de la Fournaise to a set of tabular features, we first correct the seismic signal for each day and each station to remove the instrument response.

A filter bank is then applied in between the spectral range of 0.5 - 26Hz with an initial spacing of 0.5 – 2Hz and then a spacing of 1Hz, in a similar approach to that described by Rouet-Leduc,

Hulbert, & Johnson, (2018). This results in 25 frequency bands. This frequency range is chosen as it encompasses the typical frequencies of volcanic tremor and volcano-tectonic events reported at Piton de la Fournaise **(Battaglia, Aki, & Ferrazzini, 2005; Battaglia, Aki, & Staudacher, 2005; Duputel et al., 2019)**, although eruptive signals have been reported below 1Hz more recently on broadband instruments.

We then generate features by scanning a moving window of length 1 hour across the bandpassed seismic signal, for each frequency band. The aim here is to capture distributions of the seismic data within the time windows, and relationships across different frequency bands for a given window. This has been shown to be is an effective technique to parametrize geophysical time-series data in various settings such as simulations **(Ren et al., 2019)**, laboratory experiments **(Hulbert et al., 2018; Rouet-Leduc et al., 2017; Rouet-Leduc, Hulbert, Bolton, et al., 2018)** and real Earth **(Rouet-Leduc, Hulbert, & Johnson, 2018)**. In this case, these features consist of a range of percentiles, the range of the data, as well as normalized and non-normalized higher order moments (see Supplementary Information for table describing features). For the Bory Crater seismic station (BOR), this results in 4618 non-overlapping 1 hour time windows of eruption data with 990 features generated for each window point, or 39 features per spectral band, and 38021 windows in total over the 2013-2019 period.

**2.3 Supervised Learning: Learning the Eruptive Signature of the Piton de la Fournaise**

We first use a supervised learning approach to determine the characteristics of eruptive tremor detected at the Piton de la Fournaise. In this case, the target label is the eruptive state of the volcano (erupting or dormant). For this binary classification task, we utilize a machine learning (ML) algorithm known known as gradient boosted decision trees (GBDT) **(Friedman, 2002)** with a cross-entropy loss function. We used the XGBoost implementation of GBDT to perform the modelling described in this paper, which enables parallel and distributed computation of the model **(Chen & Guestrin, 2016).**

The XGBoost model takes the features derived from a sliding time window over the seismic signal recorded at a station as inputs, and outputs a prediction on the eruptive state of the volcano. The training data in this case constitutes approximately 30% of the total data available for the BOR station, or 12000 time windows. The trained model is blind tested on the remaining 70% of the data, outputting an estimate for the eruptive state of the volcano during this period, without ever having actually seen this data. This allows us to estimate the consistency of the characteristics of the eruptive tremor throughout this period: if the spectral characteristics of the tremor are different between the training and the testing set, the model will be unable to effectively predict the eruptive state in the testing set.

Figure 2a shows the ability of our classifier to predict whether Piton de la Fournaise is undergoing an eruption or dormant based on features generated from a single time window of the continuous seismic signal recorded at the BOR station. The classifier was trained on the grey portion of the dataset (2013-2015), and tested on the violet section (2015-2019). The model performs well on the test set, with a precision of 0.99, accuracy of 0.97 and a recall score of 0.83. Figure 2b shows the receiver operating characteristic (ROC) curve for the XGBoost model, showing the true positive rate against the false positive rate. Figure 2c shows the top 10 most important features learned by the model during the training process, based on their Shapely Additive Explanations (SHAP) values **(Lundberg et al., 2018; Lundberg & Lee, 2017)**. We can see from these features that the eruptive tremor detected at the BOR station occurs consistently in the 3-5Hz frequency range. This is confirmed in Figure 2d, where we see that the 99th percentile of the 3-4Hz frequency band (red trace) of the seismic signal spikes for all eruptions, as opposed to the 14-15Hz band (blue trace).

We note that the slightly lower recall score, indicating there are a number of false negatives produced by our model in the test set. Examining Figure 2a and 2d we see that during some eruptions the level of the 3-4Hz band can drop below a given threshold, resulting in false negative prediction. Despite this, the performance of the model in detecting the state of the volcano is impressive given the small testing set consisting of only three eruptions and the fact the features are derived from the signal detected at a single station.

**2.4 Unsupervised Learning: Spectral Clustering**

Unsupervised learning is a term which describes a set of machine learning (ML) techniques used to learn relationships in datasets where no training labels are available **(Ghahramani, 2004)**. Cluster analysis, which is a subset of unsupervised learning, can be defined as the task of grouping data using similarity measures which quantify the proximity between data vectors in a given feature space **(Aminzadeh & Chatterjee, 1984)**. In this particular case, we have no explicit labelled data concerning different types or phases of eruptive behavior that Piton de la Fournaise may exhibit, or the specific seismic fingerprints of these behaviors in our feature space, so this approach is particularly well suited for our problem. In order to tackle the investigation into the seismic signature of the eruptive dynamics of Piton de la Fournaise, we turn to a clustering algorithm known as spectral clustering. Spectral clustering outperforms 'standard' clustering algorithms such as k-means in cases where the user wishes to cluster based on connectivity rather than compactness (see SI and Figure S2).

We utilize a connectivity based clustering algorithm as we hypothesize that the spectral characteristics of the seismic signal are likely to change continuously during an eruption as the system transitions between phases, thus the clusters of behavior are likely to be connected in our spectral feature space. This type of behavior in geophysical systems has previously been studied by **Holtzman et al., (2018)** who used unsupervised learning to characterize the spectral properties of seismic signals collected near geysers and identify different phases of behavior. Figures 3a and 3b show features generated using the method described in Section 2.2 on the vertical component of the continuous seismic signal recorded at the BOR station during eruptions occurring between 2013 and 2019. Here, we show two features across three different spectral bands, selected to illustrate the structure present in the 39-dimensional feature space generated for each spectral band: the variance and the minimum in a given time window for the band-passed signal. Each point in the scatterplots represents a single hour-long time window, and each color represents the relationship between the features generated for the dominant eruptive tremor band (3-4Hz) and higher frequencies. The dominant eruptive tremor band was determined using the XGBoost model discussed in Section 3.3.

The results of the spectral clustering process are shown in Figures 3c, 3d, 3e and 3f. Here we choose 6 as the optimal number of clusters, determined by using the eigengap heuristic (**von Luxburg, 2007**). For the sake of illustration we choose the 9-10Hz spectral band (the blue scatter plot shown in Figures 3a and 3b), as this is the frequency band which shows the most structure when compared to the fundamental eruptive frequency band (3-4Hz), and helps to illustrate the clustering results. We note that some clusters exhibit considerable overlap in the feature space visualization (C1, C5 and C4), there are also clusters that exhibit distinct separation in our feature space picked out by the algorithm (C0 and C2).

In order to evaluate the effectiveness of the clustering analysis on the eruptive dynamics of the Piton de la Fournaise volcano, we examine the temporal distribution of the clusters throughout the dataset. Figure 4a shows the Piton de la Fournaise eruptive state during the 2013-2019 period (red trace), and how the spectral clustering algorithm separated the features into 6 clusters (blue scatter plot). The colored bounding boxes are chosen to demonstrate the effectiveness of the combination of our feature generation method and spectral clustering in highlighting different eruptive behaviors of the Piton de la Fournaise volcano.

Figure 4b shows an irregular eruption, during which most of the features belonging to C2 occur. The eruption begins in C5, where we see the dominance of the 3-4Hz frequency band in the eruptive tremor, but also moderately elevated seismic energy in the higher frequency bands. As the eruption progresses, the eruptive state transitions to C2 such that by the end of the eruption the dominant frequency in the eruptive tremor detected at BOR is no longer 3-4Hz, but closer to the 9-10Hz range. The anomalous nature of the August 2015 eruption is further highlighted by the sudden decreases and increases in eruptive tremor amplitude towards the end of the eruption where we see consecutive transitions between C1 and C2.

A more 'usual' eruption is shown in Figure 4c, with the behavior of the volcano throughout this eruption occurring almost entirely in C1: during this type of eruptive behavior we see the main eruptive tremor frequency band (3-4Hz) dominate the seismic signal collected at BOR.

Figure 4c demonstrates the occurrence of another state transition revealed by the clustering algorithm: the eruption occurring in September 2018 initiates in C5 but transitions to C0 as the eruptive tremor amplitude increases in early October 2018. We note again that almost all time windows occurring in C0 for the whole dataset occur during this eruption.

**4 Discussion**

We note that the eruptive behavior of the Piton de la Fournaise volcano occurs in C5 during the initial phases of 7 out of 13 eruptions in our period of study (see Figure 4). Our interpretation of this result is that C5 contains time windows during which seismic signals associated with the dominant tremor band (3-5Hz) are high in amplitude (see Figure 3) but also moderate amplitudes in the higher frequency ranges, which may result from rockfall or volcano-tectonic events. It is also worth noting broadband seismic signals can be related to impulsive events related to the migration of dykes and brittle failure within the edifice of the volcano **(Duputel et al., 2019),** but also subsidence of the caldera flow and lava tube collapse related to the lava load.

The anomalous signal recorded during the eruption of August-October 2015 can thus be interpreted in a similar manner: the sustained high frequencies belonging to C2, which occur almost entirely during the aforementioned eruption, are likely due to the subsidence of the caldera floor or lava tube collapses due to the load of the lava flow from the eruption. Indeed, this eruption resulted in a large total effusive volume of $35.5 \times 10^6 m^3$, the largest amount recorded for the 2013-2018 period. Geodetic data indicates the volcanic edifice continually inflated until the end of September, indicating pressurization of the shallow magma reservoir as a consequence of the impulsive ascent of deeper magma. Furthermore, **Coppola et al., (2017)** reported a magnesium-rich magmatic source becoming evident in correspondence with this new period of inflation, with an increase in $SO_2$ emissions and strong $CO_2$ enrichment in summit fumaroles during this phase of the eruption, providing further that the shallow reservoir was refilled by deep magma, leading to the re-pressurization of the magmatic plumbing system. This re-pressurization is likely responsible for the sustained flow of lava throughout the long duration of this eruption, inducing subsidence and lava tube collapse and the associated anomalous seismic signals observed during the second phase of this eruption. We note furthermore that the fissures and lava flow occurred in relatively flat areas of the volcano, reducing the flow of lava and inducing stacking and increasing the lava load in the area.

Figure 5 shows the unique position of this eruption in our cumulative feature space. To produce this figure we calculate the features for the vertical components of three broadband stations (FJS, CSS and FOR) that are approximately equidistant from the Dolomieu crater **(De Plaen et al., 2016)** (see Figure 1). We average the feature values for time windows at which data is available across all three stations, and sum these across the duration of the respective eruptions, to give a sense of the feature values coupled with the overall duration of the eruptions. For each eruption we impose the condition that there needs to be over 50% of the duration of the eruption available in terms of the features. From Figure 5 we can see that features associated with the dominant tremor band as well as higher frequencies were consistently high throughout the duration of this eruption, even when compared to similar eruptions of longer duration (> 1 month) during which re-pressurization was observed, such as the April 27 – June 1 2018 eruption. We also note C0 represents another eruptive transition picked out by the clustering algorithm: according to OVPF reports concerning the September-November 2018 eruption, gas pistons were recorded from October $3^{rd}$ onwards, marking a transition in the degassing of the eruption and accompanying the sharp increase in eruptive tremor recorded. The increase of tremor recorded, which corresponds to the C0 cluster is likely related to the gradual closing of the eruptive cone, leading to a higher amplitude of 'resonance' **(B. Chouet, 1988)** thus increasing the tremor amplitude. We note that following the closing of the eruptive cone, the eruption continued mostly in lava tubes, which is likely the source of the higher frequencies observed for C0.

## 5 Conclusions

We have shown that both supervised and unsupervised machine learning techniques provide effective ways to probe volcanic behavior. Using XGBoost we established the dominant eruptive tremor frequency at the Piton de la Fournaise volcano. By utilizing spectral clustering on features derived from the seismic signal recorded at the BOR seismic station we show that we can uncover eruptive dynamics in the generated feature space such as the August 2015 eruption during which a relatively large sustained lava flow was observed, and the closing of the eruptive vent during the September 2018 eruption.

## Acknowledgments

C. R., B. R. L., P. J acknowledge institutional support (LDRD) at Los Alamos and the US DOE Office of Science, Geoscience, for funding. We thank Ian McBrearty, James Theiler, Andrew Delorey and Daniel Trugman for useful discussions. The data were obtained from the the Observatoire Volcanologique du Piton de la Fournaise (OVPF) network.

## Data Availability

The data used is publicly available and can be found online. The seismic data used for this study were obtained from the Observatoire Volcanologique du Piton de la Fournaise (OVPF) network (https://www.fdsn.org/networks/detail/PF/)and can be downloaded from the IPGP Data Center

(http://ws.ipgp.fr/fdsnws/dataselect/1/), the IRIS Data Management Center (IRISDMC, http://service.iris.edu/fdsnws/dataselect/1/) or RESIF (http://ws.resif.fr/fdsnws/dataselect/1/)

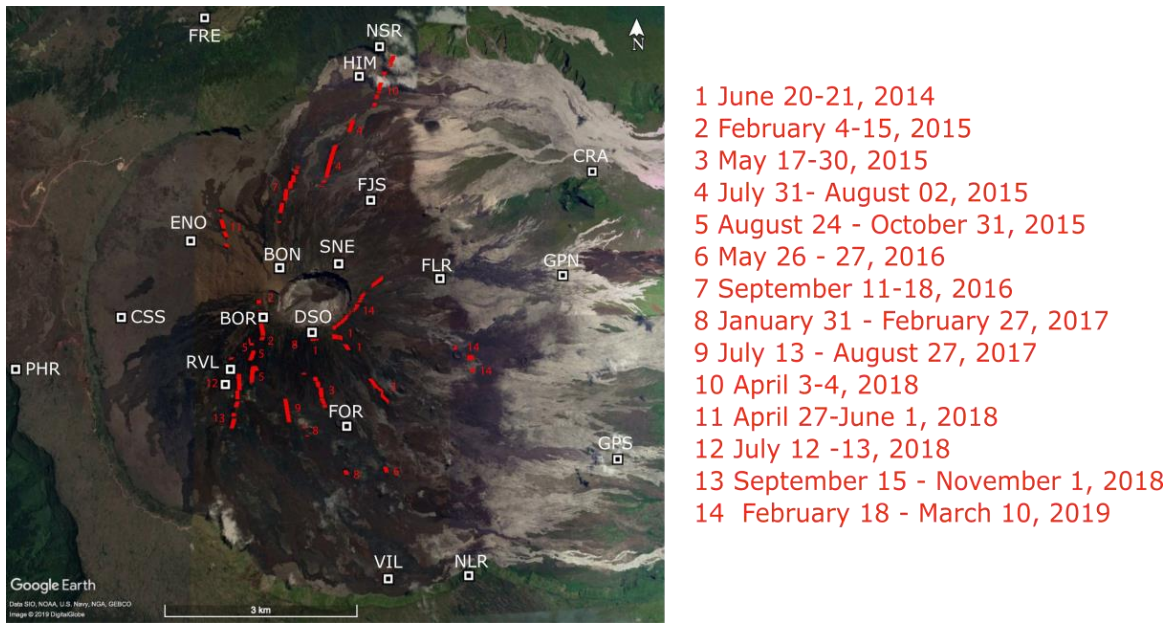

**Figure 1**. Map of the seismic stations located near the Piton de la Fournaise volcano, with the locations of fissures resulting from eruptions highlighted and numbered in red.

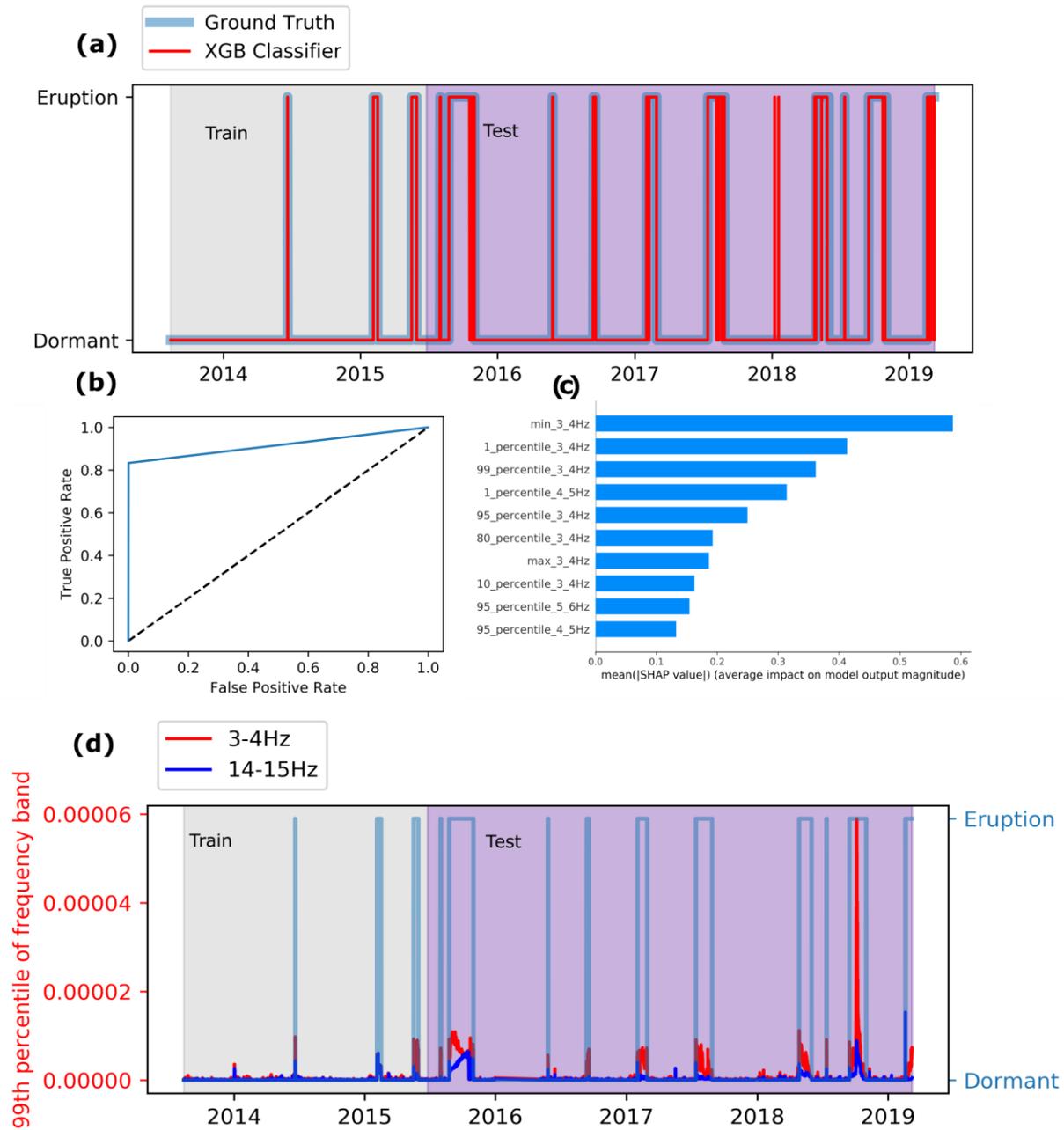

**Figure 2:** (a) Performance of the XGBoost classifier trained to predict the eruptive state of Piton de la Fournaise volcano based on features derived from the continuous seismic signal recorded at the BOR station. The eruptive state (target) is shown in blue, and the model predictions in red. The model is trained only on the grey portion of data. (b) Receiver operating characteristic (ROC) curve for the XGBoost model, showing the true positive rate against the false positive rate. (c) Top 10 informative features derived from the continuous seismic signal, based on SHAP importance value, demonstrating that the volcanic tremor signal resides mostly in the 3-5Hz band. (d) 99th percentile of the 3-4Hz (red) and 14-15Hz (blue) plotted with eruptive state. The 3-4Hz band value increases beyond a consistent threshold for eruptive states across the data set, allowing the machine-learning algorithm to recognize the occurrence of eruptive tremor.

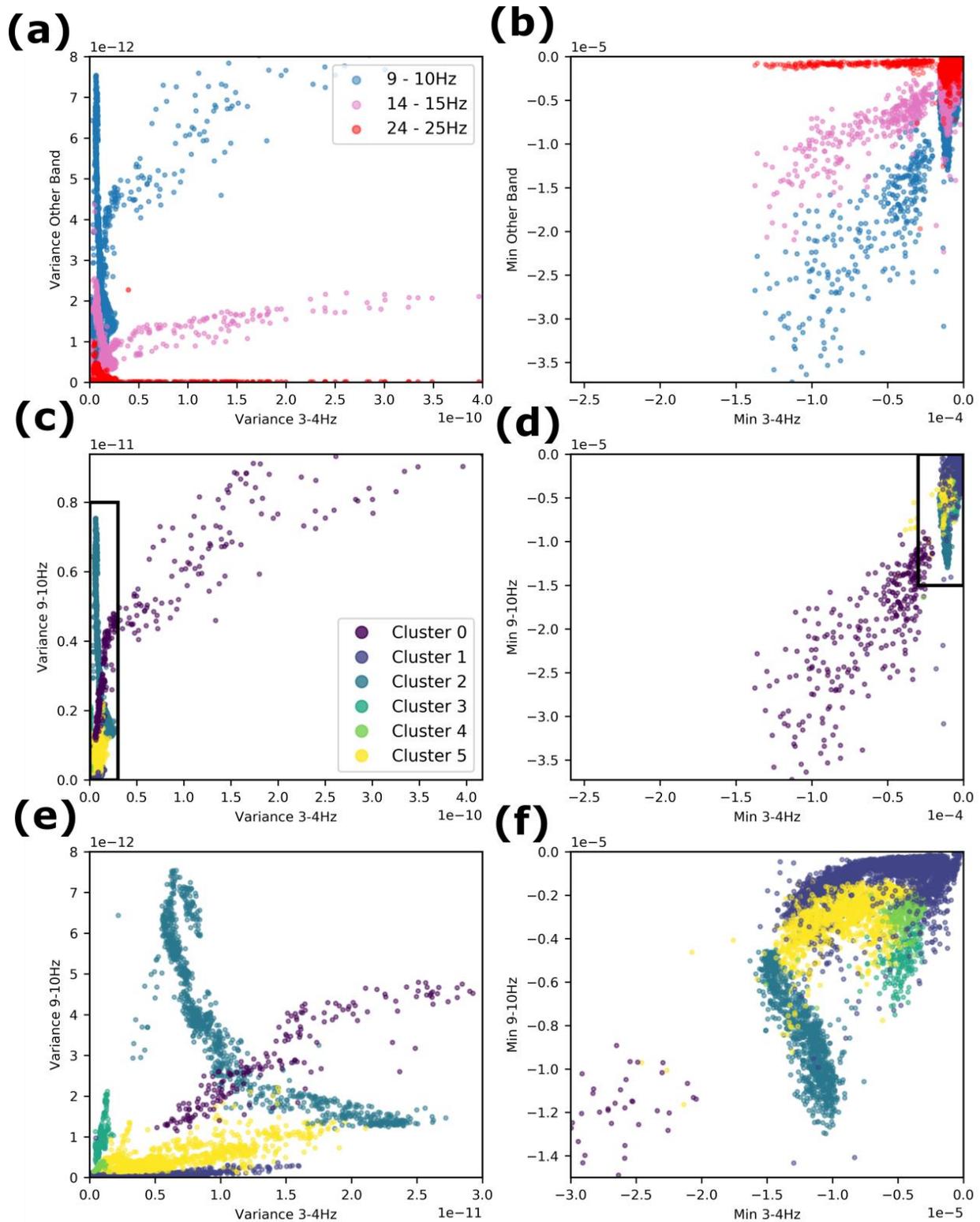

**Figure 3:** Spectral relationship between features generated at higher frequencies (9-10hz, 14-15Hz and 24-25Hz) and the dominant volcanic tremor frequency (3-4Hz). The features shown here are the (a) variance and (b) the minimum value for hour-long time windows of the band-

passed continuous seismic signal for the station BOR in the aforementioned spectral bands. We show the results of the spectral clustering on these features in figures (c) and (d) respectively with the dominant tremor band plotted against the 9-10Hz band. The black bounding boxes are shown in greater detail in (e) and (f) for the spectral band variance and minimum value, demonstrating the structure present amongst the features as picked out by the clustering algorithm.

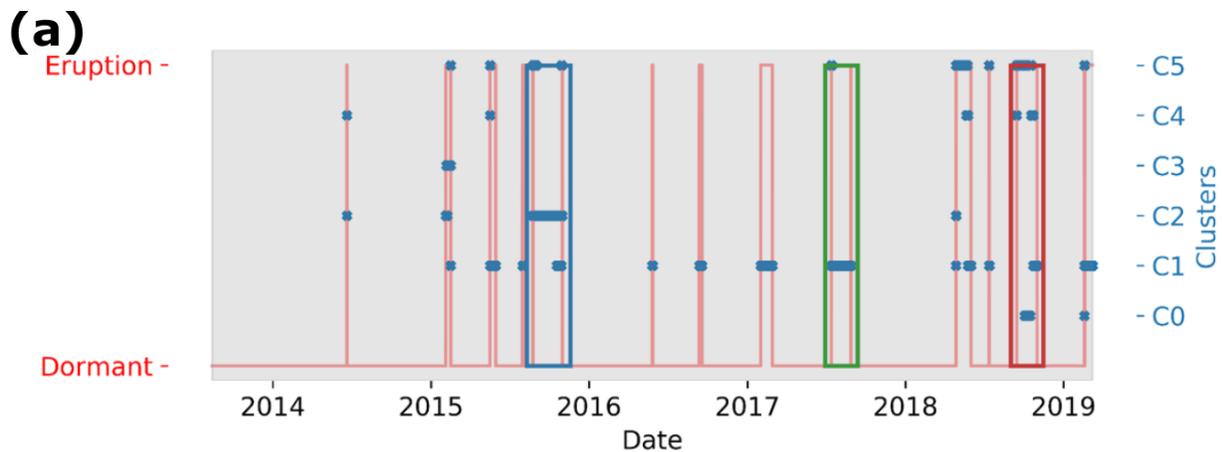
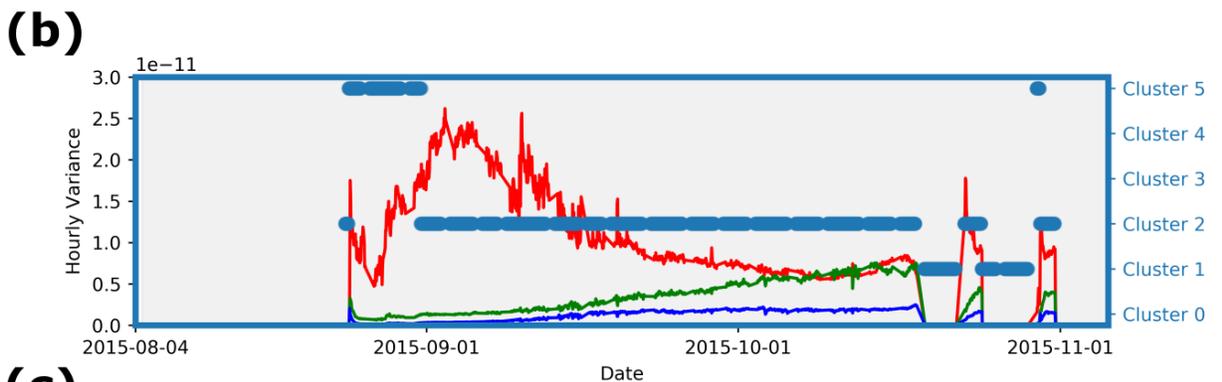
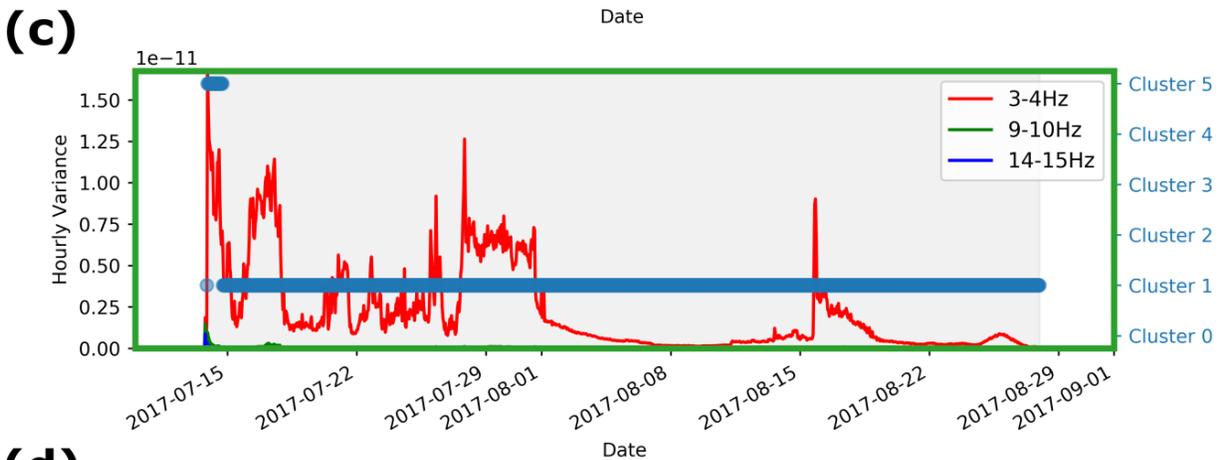
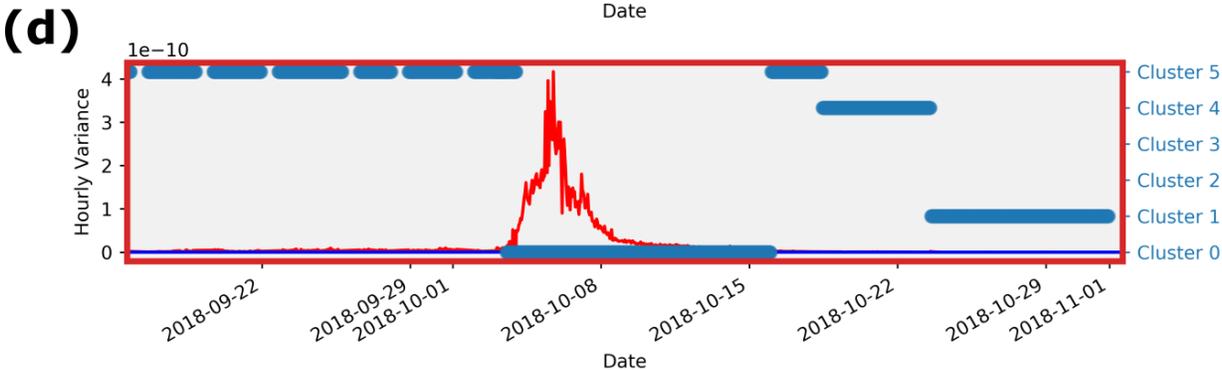

**Figure 4:**

Temporal distribution of the clusters shown in Figure 3. **(a)** The temporal distribution for the entire dataset (2013 - 2019). The three bounding boxes are chosen as representative examples of the clusters. **(b)** An 'anomalous' eruption, exhibiting a transition between Cluster 5 and Cluster 2: during this eruption we see an initial dominant frequency range for the eruptive tremor between 3-4Hz, but as the eruption progresses higher frequencies start to dominate the eruptive tremor for the BOR station. We also note the abrupt decrease in amplitude of the eruptive signals in the waning stages of the eruption, where the eruptive signals detected at BOR transition between Clusters 1 and 2 repeatedly. **(c)** A more 'regular' eruption residing mostly in Cluster 1 (see Figure 3e) where the variance of the 3-4Hz frequency band of the eruptive tremor tends to be strong relative to higher frequencies for the BOR station. **(d)** An eruption time windows located in cluster 0. We note that both the dominant frequency band *and* the higher frequencies are strong in this cluster.

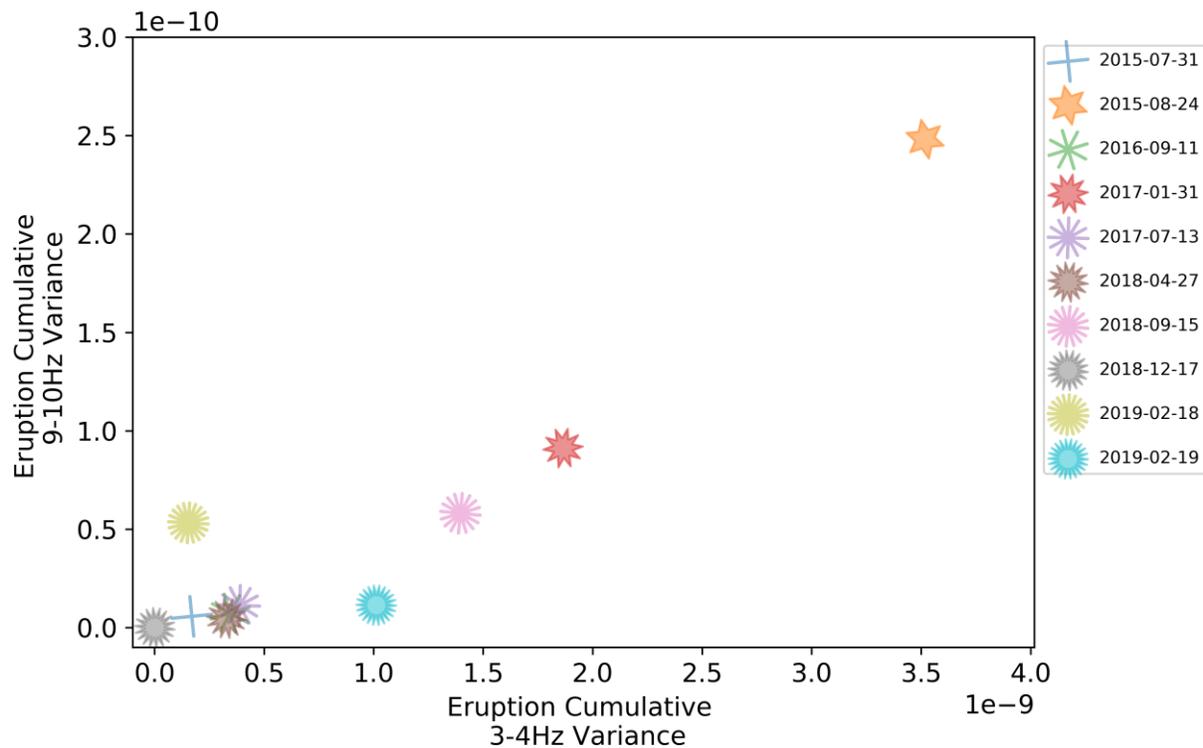

**Figure 5:** Cumulative variances for the 3-4Hz band and the 9-10Hz band, averaged for stations CSS, FOR and FJS (see Figure 1) which are approximately equidistant from the Dolomieu crater, confirming the 'anomalous' nature of the August 2015 eruption in terms of the seismic signals

detected throughout the duration of the eruption. We have removed eruptions for which more than 50% of the time windows were missing across the three stations combined.

Manuscript submitted to *Geophysical Research Letters*Ort, M. H., Di Muro, A., Michon, L., & Bachèlery, P. (2016). Explosive eruptions from the interaction of magmatic and hydrothermal systems during flank extension: the Bellecombe Tephra of Piton de La Fournaise (La Réunion Island). *Bulletin of Volcanology*, *78*(1), 5.

Peltier, A., Bachèlery, P., & Staudacher, T. (2009). Magma transport and storage at Piton de La Fournaise (La Réunion) between 1972 and 2007: A review of geophysical and geochemical data. *Journal of Volcanology and Geothermal Research*, *184*(1–2), 93–108.

Plaen, R. S. M. D., Lecocq, T., Caudron, C., Ferrazzini, V., & Francis, O. (2016). Single-station monitoring of volcanoes using seismic ambient noise. *Geophysical Research Letters*, *43*(16), 8511–8518. https://doi.org/10.1002/2016GL070078

Ren, C. X., Dorostkar, O., Rouet-Leduc, B., Hulbert, C., Strebel, D., Guyer, R. A., et al. (2019). Machine Learning Reveals the State of Intermittent Frictional Dynamics in a Sheared Granular Fault. *Geophysical Research Letters*. https://doi.org/10.1029/2019GL082706

Ross, Z. E., Yue, Y., Meier, M.-A., Hauksson, E., & Heaton, T. H. (2019). PhaseLink: A Deep Learning Approach to Seismic Phase Association. *Journal of Geophysical Research: Solid Earth*, *124*(1), 856–869. https://doi.org/10.1029/2018JB016674

Rouet-Leduc, B., Hulbert, C., Lubbers, N., Barros, K., Humphreys, C. J., & Johnson, P. A. (2017). Machine Learning Predicts Laboratory Earthquakes. *Geophysical Research Letters*. https://doi.org/10.1002/2017GL074677

Rouet-Leduc, B., Hulbert, C., & Johnson, P. A. (2018). Continuous chatter of the Cascadia subduction zone revealed by machine learning. *Nature Geoscience*. https://doi.org/10.1038/s41561-018-0274-6